# Validating Warehouse Picking Strategies Using Simulation: Case Study of a Plumbing Equipment Firm

Phattara Khumprom[1*], Wanatchapong Kongkaew[2], Antoun Yaacoub[3],

Nattakit Thanawitsatien[4]

*Graduate School of Management and Innovation (GMI), King Mongkut's University of Technology Thonburi (KMUTT)*
*126 Pracha Uthit Rd, Bang Mot, Thung Khru, Bangkok 10140, Thailand*

*Industrial and Manufacturing Engineering, Faculty of Engineering, Prince of Songkla University (PSU)*
*15 Kanchanavanit Road, Hat Yai, 90110 Songkhla, Thailand*

*aivancity School for Technology, Business & Society*
*151 Bd Maxime Gorki, 94800 Villejuif, France*

[1]phattara.khum@kmutt.ac.th
[2]wanatchapong.k@psu.ac.th
[3]yaacoub@aivancity.ai
[4]nattakit.than@kmutt.ac.th

***Abstract*— In today's competitive business environment, efficient logistics are essential, especially in industries where timely delivery matters. This research aims to improve warehouse picking cycle time through simulation-based analysis, using a leading plumbing equipment distributor in Thailand as a case study. The study identifies inefficiencies such as disorganized storage and poor placement of high-frequency items that slow down picking. To address this, an optimized storage approach using ABC analysis is proposed, prioritizing high-demand items near the entrance. Three storage policies—Fixed, Random, and Combination (Fixed Zone)—are tested with a Zone Picking strategy through simulation to identify the most efficient picking routes. The findings provide insights for improving warehouse layout and inventory placement to enhance overall performance.**

## I. Introduction

In today's highly competitive industrial landscape, efficient logistics are crucial to meeting customer demands and ensuring timely delivery. Warehousing plays a key role in this process, with activities such as put-away, picking, and goods transfer directly affecting performance. However, many industries face inefficiencies in these operations. This study focuses on a leading plumbing equipment manufacturer in Thailand, whose warehouse suffers from disorganized storage and poor placement of frequently picked items, leading to longer picking times and reduced efficiency. To address this, the research compares the current storage system (As-Is) with an optimized approach using ABC analysis to prioritize high-demand items near the entrance. Three storage policies—Fixed, Random, and Combination (Fixed Zone)—are evaluated under a Zone Picking method through warehouse simulation to identify the most efficient picking routes and improve overall warehouse performance.

## II. Literature Review

Most studies on warehouse operations have primarily focused on two areas: order picker routing and product placement strategies. Order picker routing, which involves finding the shortest path for pickers within a warehouse, is a complex variation of the NP-hard Traveling Salesman Problem [1]. In contrast, strategies for determining optimal product locations to support efficient order picking have received relatively limited attention. One of the earliest contributions in this area came from Jarvis and McDowell [2] proposed a model aimed at minimizing average picking time through strategic product placement. After that, Jewkes, Lee, and Vickson [3] explored product positioning along a picking line, where orders were fulfilled by moving a container past storage locations and loading the required quantities. Notably, no previous research, to our knowledge, has examined product location in conveyor belt-based systems like the one studied in this paper.

Simulation, meanwhile, has long been a widely used method in warehouse research. From early studies emphasizing its value [4], [5] to more recent applications, simulation has largely served to test the effectiveness of warehouse layouts and handling systems. Many of the operational research (OR) models discussed earlier have been validated using simulation. Hybrid methods combining optimization and simulation, such as those by Gue et al. [6] and Renaud and Ruiz [7], exemplify this approach. Despite the abundance of studies in the field, however, broad theoretical contributions remain limited. This is mainly because optimal product placement depends heavily on specific warehouse layouts, picking strategies, and technologies, leading most research to focus on case-specific applications.

## III. The Warehouse Operations

This research aims to explore and compare the efficiency of different storage policies using the zone picking method, supported by a warehouse simulation model. The study also seeks to identify effective strategies for designing storage layouts and determining optimal product placement to enhance overall warehouse efficiency.

### *A. Overall information of the case studies' products and warehouse*

The case study examines a Thailand-based plumbing equipment manufacturer with domestic and international operations. Its 1,740 $m^2$ warehouse stores 988 pallets across five zones—Office, Special Order, Packing, Storage, and Dispatch—and uses three- and four-tier selective racking systems to optimize space, accessibility, and workflow efficiency.

### *B. Case studies' warehouse operation*

This study maps warehouse operations, collects process data, and uses simulation to identify key factors affecting cycle time, with the goal of determining the optimal picking strategy. Figures 1–3 present the warehouse layout, shelving units, and process times.

The warehouse operates through eight main processes:

*1) Transportation* (07:00–08:30): Pallets are moved from the staging area to delivery vehicles and loaded in LIFO order for efficient unloading.

*2) Receiving* (08:00–09:00, 11:00–12:00, 15:30–16:00): Goods from production are inspected, documented, and recorded in the ERP system.

*3) Transfer-In*: Pallets are moved via elevator to the second floor and placed in designated storage or queue lanes.

*4) Put-Away*: Tagged pallets are stored using electric stackers following FIFO rules.

5) Picking (09:00–11:00): Staff retrieve items based on daily orders, using hand pallet trucks or stackers, and move them to the Special Area.

*6) Sorting* (10:30–11:00, 13:00–15:30): Picked items are grouped by customer order—by route for Bangkok deliveries or for repackaging for upcountry customers.

*7) Packaging* (10:30–11:00, 13:00–15:30): Sorted goods are repacked, labeled, and prepared for shipment.

*8) Transfer-Out* (16:00–17:00): Packed pallets are moved to the first-floor staging area for dispatch.

Data analysis identified key parameters influencing cycle time, revealing that put-away and picking consume the most time and resources. Optimizing these two processes can significantly improve overall warehouse efficiency. More in-dept inventory flow process from the case study as described in Figure 5.

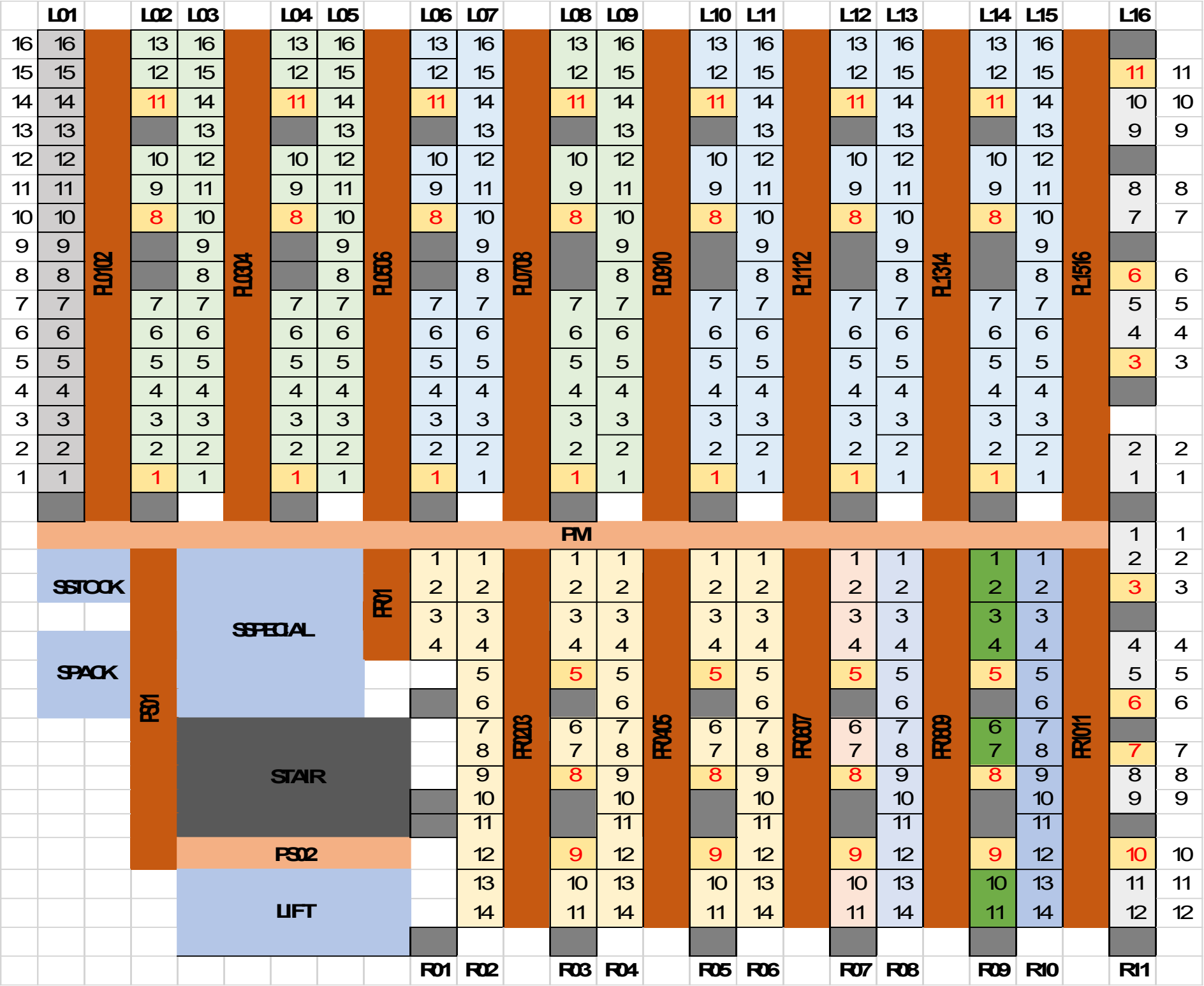

Fig. 1  Current warehouse's floor plan.

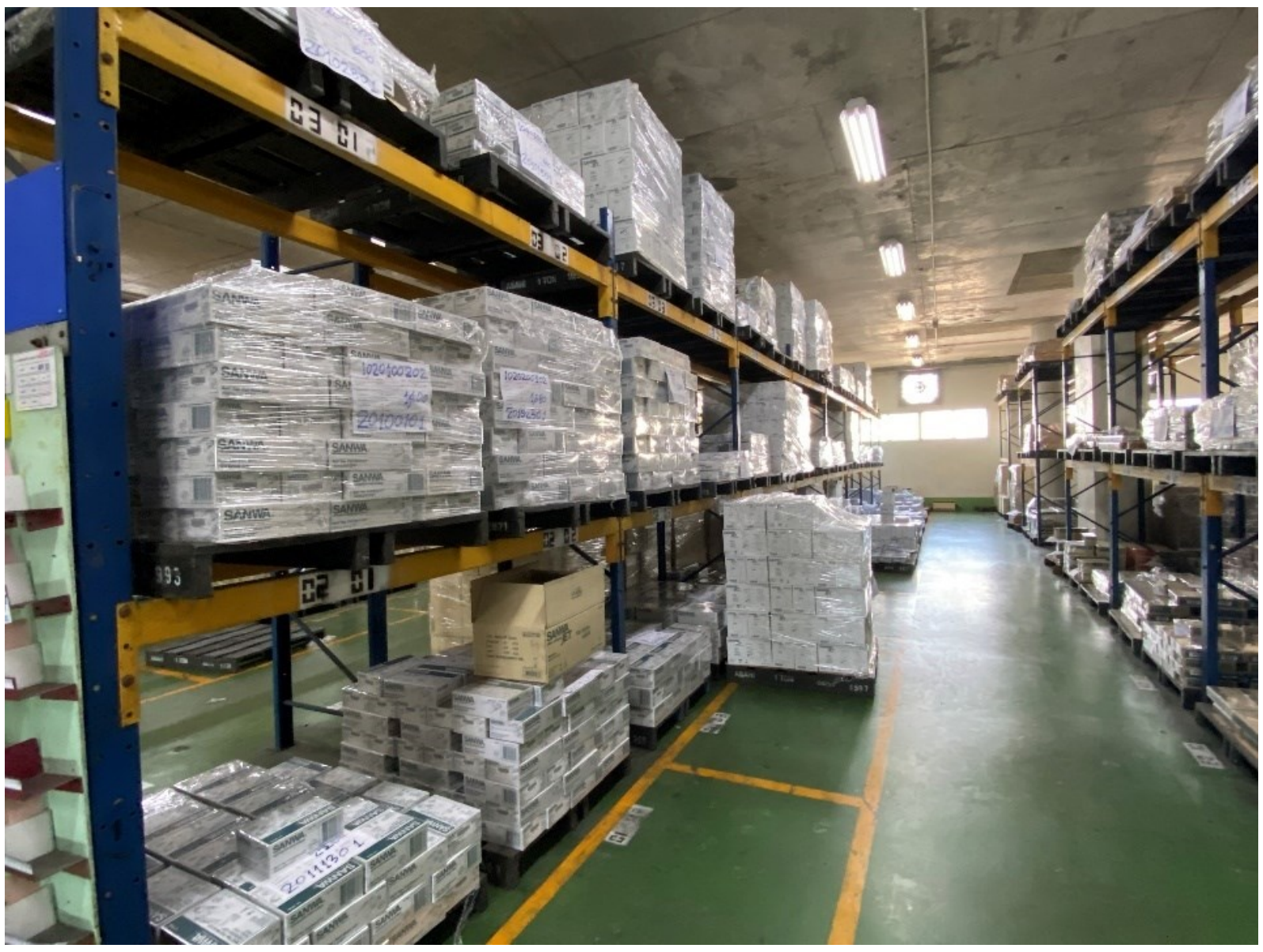

Fig. 2  Current shelves arrangement image from the site.

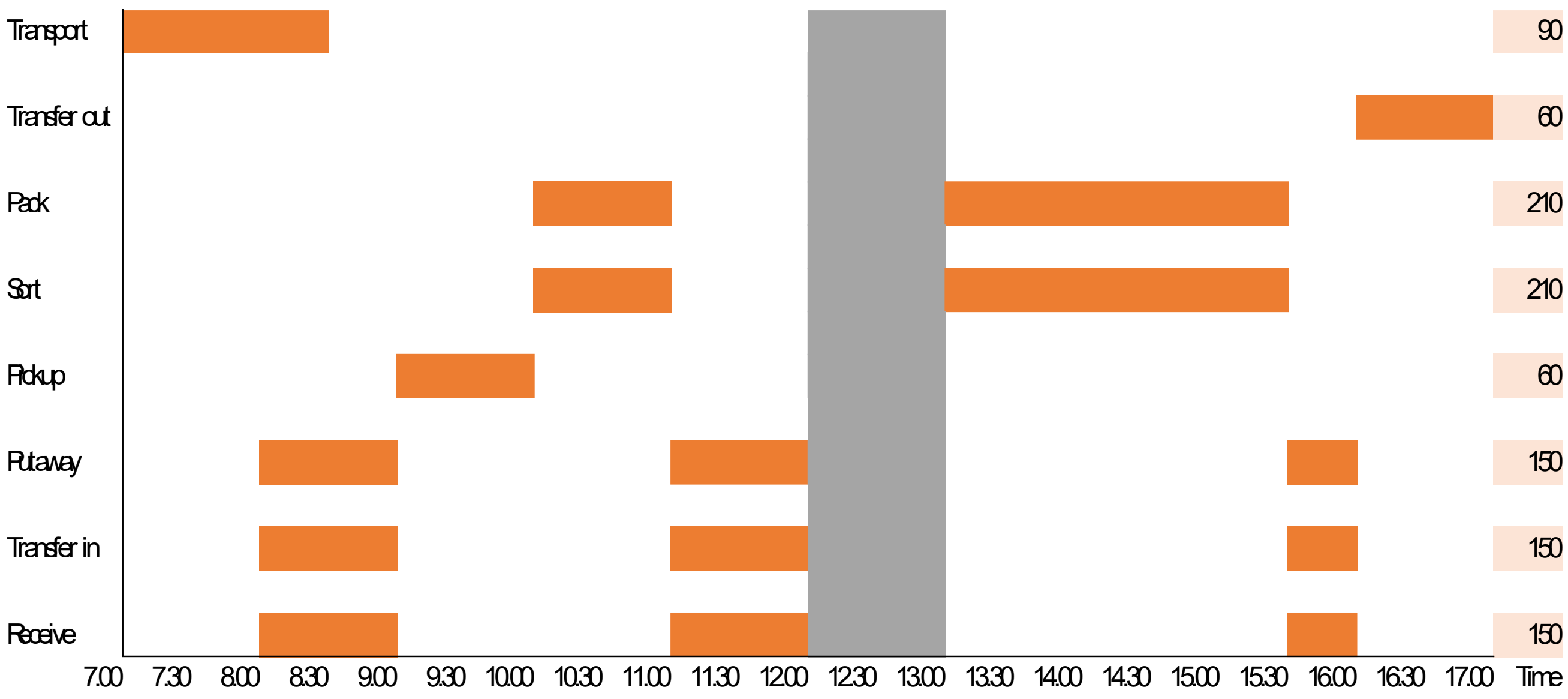

Fig 3. Operation time/period of picking procedure in the case studies' warehouse.

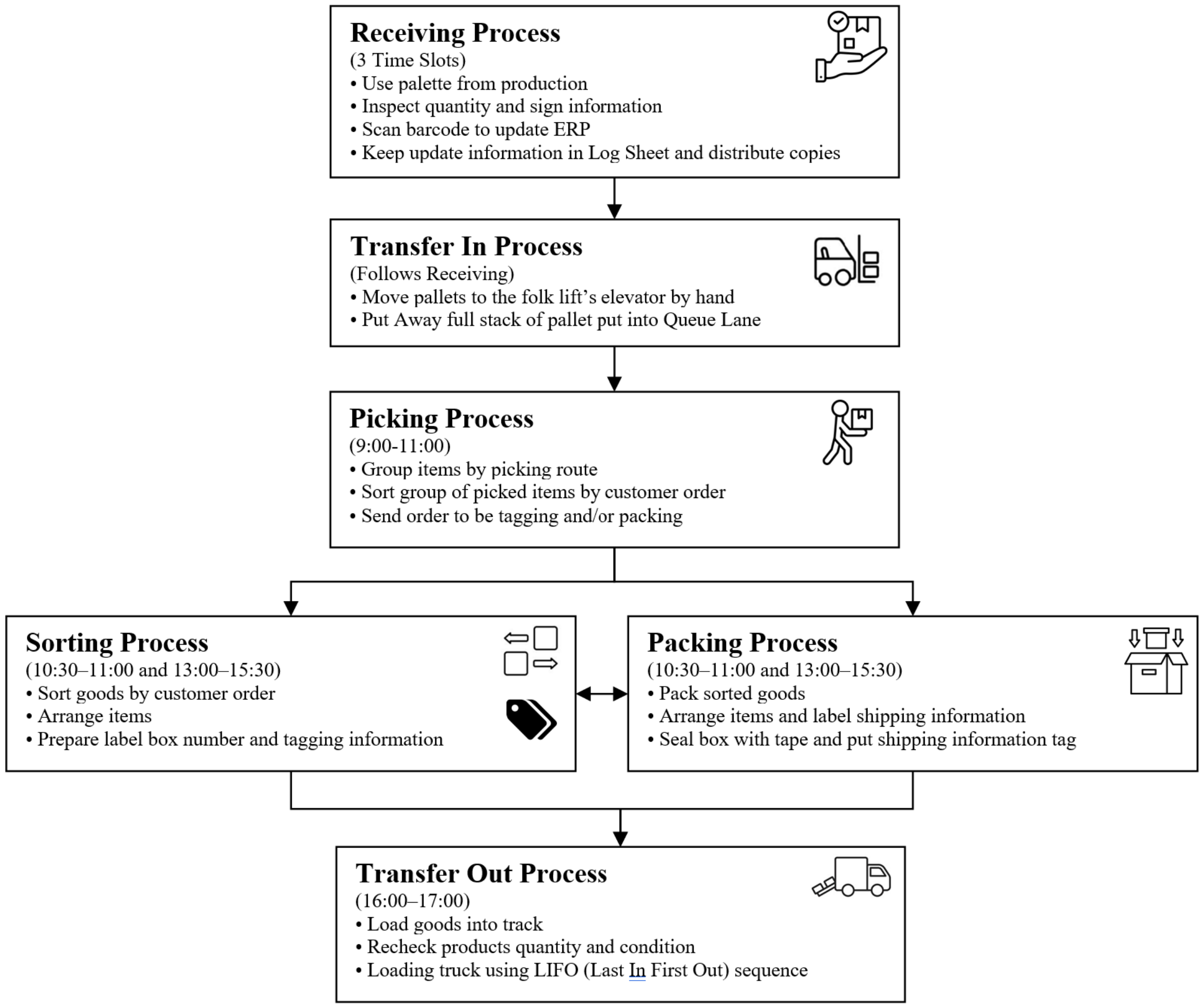

Fig. 4 Overall warehouse's operation procedures.

## IV. Modeling the Picking Activities

The simulation uses customer orders and system parameters such as storage locations, picker speed, and replenishment cycle. Orders are grouped by delivery truck, arranged in reverse delivery sequence, and sorted by shelf proximity. Each order is processed sequentially through two main activities: walking (constant time) and handling (based on quantity picked). If stock is sufficient, total picking time equals walking plus handling time, after which inventory is updated. In case of shortages, the picker's arrival time at the stock-out point is estimated, and items are either fully picked after replenishment or partially processed until restocking. The process was validated by the warehouse operations manager to ensure accuracy, with model details presented in the next section.

### *C. The picking process*

#### *1) Fundamental Picking Methods*

The basic picking method which conducted from the case study can generally be categorized into three approaches:

- The picker physically travels on foot to the storage location.
- The picker utilizes a vehicle to access the storage location.
- The items are transported from storage to the picker's designated work area.

#### *2) Order-Picking Management Systems*

Based on the case study this system may be classified into four principal models:

- Area System – The picker receives an order and proceeds to the designated storage area to retrieve the required items. Subsequently, the items are transferred to the packing area prior to shipment to the customer.
- Zoning System – The storage facility is divided into zones, often defined by aisles or sections. Each picker is assigned responsibility for a specific zone. Customer orders are divided accordingly, and items are retrieved within their respective zones and then consolidated in a designated order assembly area.
- Sequential System – Similar to the zoning system, but with a sequential workflow. After items are retrieved from one zone, the order is transferred to the next zone for further picking. This process continues until the order has been fully completed.
- Multiple Order System – Multiple customer orders are grouped and aggregated into batches. Item quantities are consolidated, and pickers retrieve bulk quantities from their designated zones. The items are subsequently transported to a sorting area, where they are separated and organized according to individual customer orders.

#### *3) Order-Picking Route Patterns*

The order picking pattern from the case study can generally be classified into two approaches as follows.

- Non-Routing Pattern – In this approach, pickers independently determine their own travel paths when retrieving items. Since the routes are not predetermined or standardized, this method is seldom adopted in practice due to its tendency to create inefficiencies in both time utilization and operational costs.
- Sequential Order-Pick Pattern – This method assigns numerical sequences to storage locations along each aisle, thereby establishing a structured order for item retrieval. The sequence can be organized in various ways, depending on whether the picking strategy involves single-side picking or multi-side picking, in accordance with the warehouse layout and storage system design.

More in-dept of picking process flow from the case study as illustrated in Figure 6.

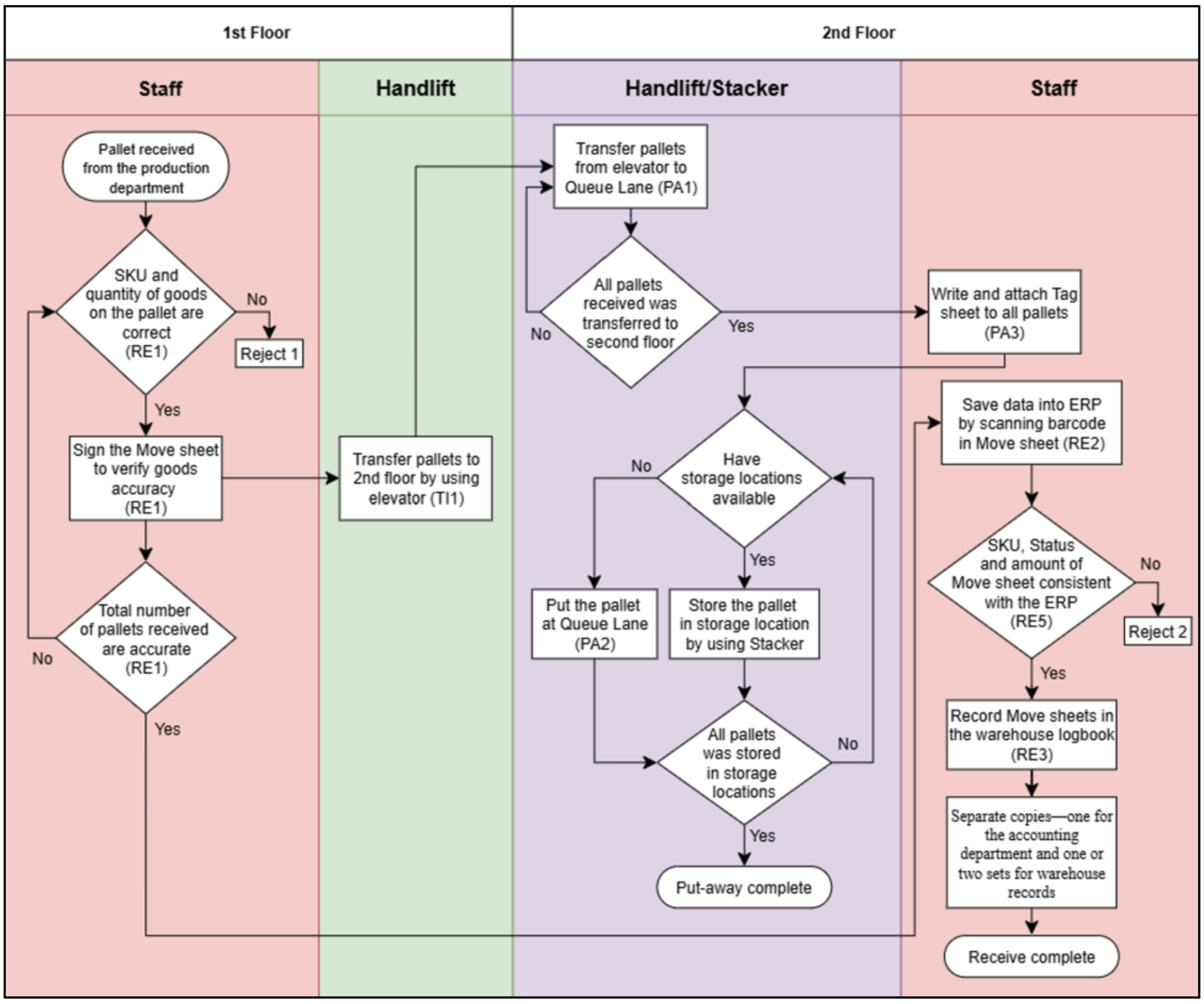

Fig. 5 Process flow diagram of goods receiving and the process of transferring goods into the warehouse.

## V. Implementing the Model

### *A. General information of the simulated model*

The simulator replicates real system behavior by accurately modeling the operational decision processes. Built on a discrete event simulation (DES) engine using the three-phase approach [8], it represents system changes as instantaneous events over time [9]. The simulator consists of three main components: a database, an event list, and a simulation clock that controls simulated time progression. The data collected for the purpose of analyzing and simulating warehouse operations can be summarized as follows.

*1) Warehouse Layout Data:* Includes the warehouse structure, rack arrangement, and distances to entry/exit points, forming the basis for simulating operations.

*2) Order and Receiving Data:* Covers order records and inbound shipments from the peak month, reflecting the company's operational characteristics.

The database contains orders and system parameters, currently using historical data, with future pseudo-random data planned for robustness testing. The simulation engine consists of an event list and a clock that advances simulated time. Events are processed chronologically, generating future events as needed, then removed from the list as the clock moves to the next event.

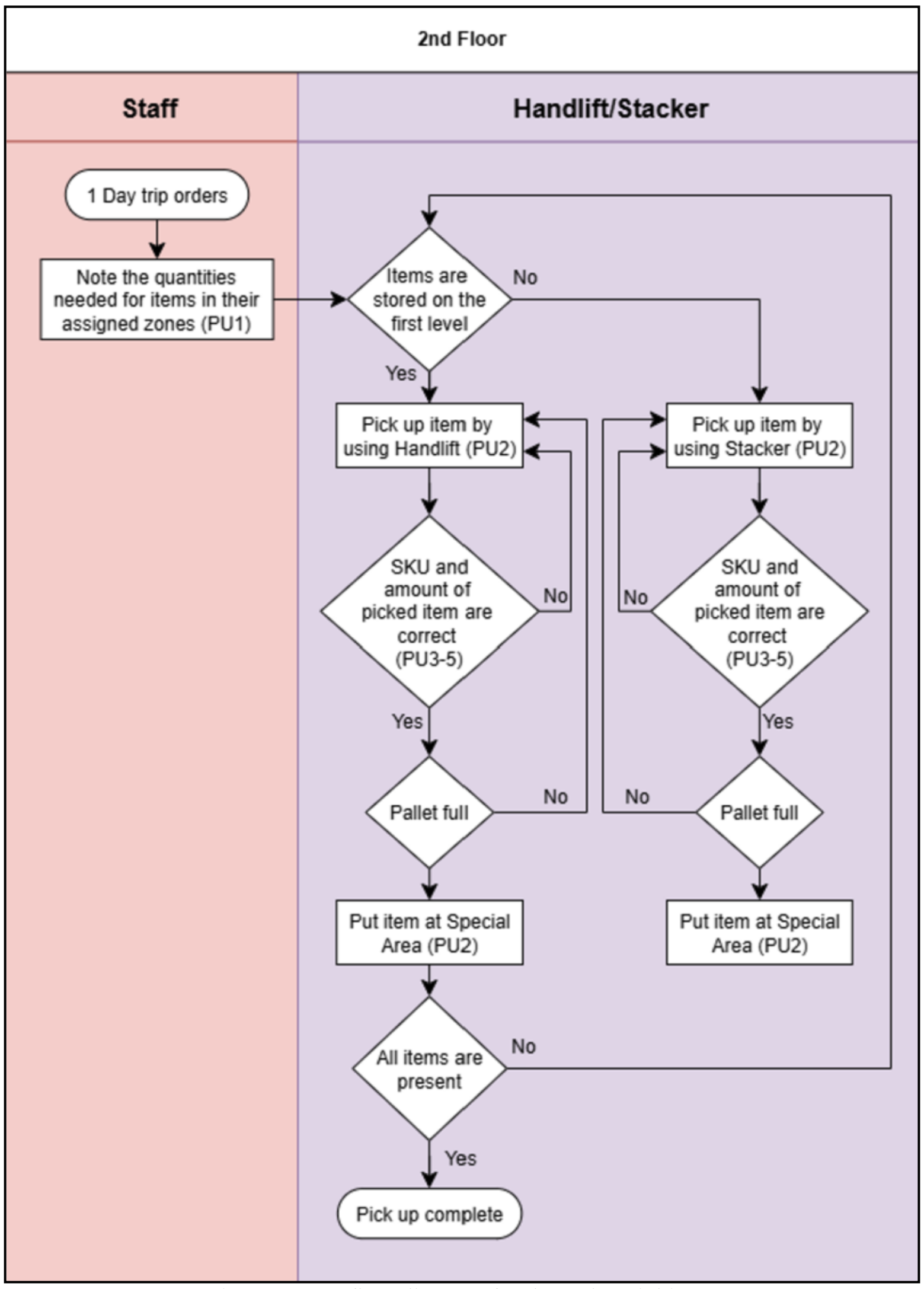

Fig.6 process flow diagram for the order-picking process.

Warehousing operations are modeled with three event types. Initially, the event list contains two entries: the first order picking and the first replenishment. The simulation clock, starting at zero, advances to and executes the earliest event.

For example, suppose the first event is $SPO_1$ (start picking order 1) at time $t_1$. Executing $SPO_1$ involves checking the stock for all lines in the first order. If every line can be fulfilled, an $SPO_2$ (start picking order 1) event is scheduled for $t = t_1$ + walking time + handling time, with handling time calculated as described previously. $SPO_1$ is then removed from the list, and the clock advances to the next event (either $RP_1$ or $SPO_2$).

If a stock-out is detected during the inventory check, a PP (partial picking) event is created at the estimated time the picker reaches the stock-out location (as explained earlier) instead of an SPO. When a PP event is executed, the inventory at that location is checked again. If the demand can be met, either an SPO—if the remaining order lines can be completed—or another PP for the next stock-out location is scheduled. If the demand still cannot be met, another PP is scheduled for the same location immediately after the next replenishment event.

This approach speeds up simulation by reducing event list operations when orders are fulfilled without stock-outs, though it approximates the picker's arrival time at stock-out locations.

When a replenishment event is executed, it schedules the next replenishment for $t$ = *current time* + $t_r$, where $t_r$ is drawn from a normal distribution fitted to historical replenishment times observed in practice. In generating a new replenishment event, it is also necessary to determine which product will be restocked. Various selection strategies can be applied, depending on the information used and the desired degree of synchronization between picking and replenishment. For instance, a purely random rule could select, with equal probability, any storage location eligible for refilling—meaning it has enough space for a full pallet. However, this approach offers no synchronization between picking and replenishment and was discarded after discussions with replenishment staff. Instead, the implemented rule, chosen for its simplicity and intuitive appeal, prioritizes replenishing the product with the lowest inventory level.

The database stores all simulated scenarios, including results, performance metrics, and experiment parameters (e.g., replications, simulation length) for later analysis.

The simulation was coded in Python using Visual Studio Code and partially Jupyter Notebook, developed from scratch based on a Thai plumbing equipment firm case study. Object-oriented code was debugged and validated step by step.

Python was chosen for its flexibility in testing multiple location and replenishment strategies, ease of dynamic modifications, and strong library support. It was preferred over other languages (C++/Java) and simulation software (ARENA/FlexSim) and aligned with the company partner's existing development tools.

*B. The input data for the simulation*

*1) Model Settings*: Includes simulation dates, working hours, equipment configuration (position, quantity, speed, turning time, operators, activities), process times (pick, put, move between layers, plus handling per pallet/item), activity plans and durations (Pick-up, Putaway, Sorting, Transfer), and miscellaneous parameters (max pallet weight/volume, transfer intervals, sorting time per SKU, operator workload, and breaks).

*2) Item Details* (ItemDB): Item code, category, weight, storage zone, and quantity per pallet.

*3) Storage Rack Locations* (AssetLocation): Rack/aisle IDs, layer/slot sequences, coordinates (X/Y/Z), dimensions, parent node info, and direction.

*4) Item Location & Quantity*: Rack coordinates, zone, item code, quantity, and manufacturing date.

*5) Customer Orders* (June 2020): Order date/time, order number, item code, quantity, and total weight.

*6) Inbound Goods* (June 2020): Putaway date/time, order number, item code, quantity, total weight, and manufacturing date.

*C. Flow process of the simulation models*

The warehouse simulation model supports three storage strategies:

*1) Fixed Location*: Each item has a permanent storage slot. Upon receipt, the system assigns the nearest available slot for that item, dispatches a stacker to place it, and records its details. If no slot is available, the item is added to a waiting list.

*2) Randomized Storage*: Items can be stored in any available slot. The model uses Fixed Zone mode to simulate this, assigning items to the nearest vacant slot, dispatching a stacker, and recording the item. Unavailable slots are added to a waiting list.

*3) Fixed Zone*: Only the storage zone is predetermined; items can occupy any slot within the zone. The system assigns the nearest slot within the zone, dispatches a stacker, and records item details.If no additional inbound items remain, the stacker returns to the Special Area or proceeds to other tasks. In cases where no storage slots are available within the designated zone, the item is recorded in the waiting order list, and the system moves on to the next inbound item.

The current storage strategy employed by the case study company is a combination of Fixed Location and Randomized storage, commonly referred to as Fixed Zone storage. In this approach, storage zones (Zone) are assigned to specific product categories using fixed-location principles, while within each zone, items can be stored in any available slot (Randomized) according to the designated zone. The order-picking strategy follows a daily cumulative approach, where items are picked based on the total daily demand for each product category (One Day Trip / Pick-up by Zone).

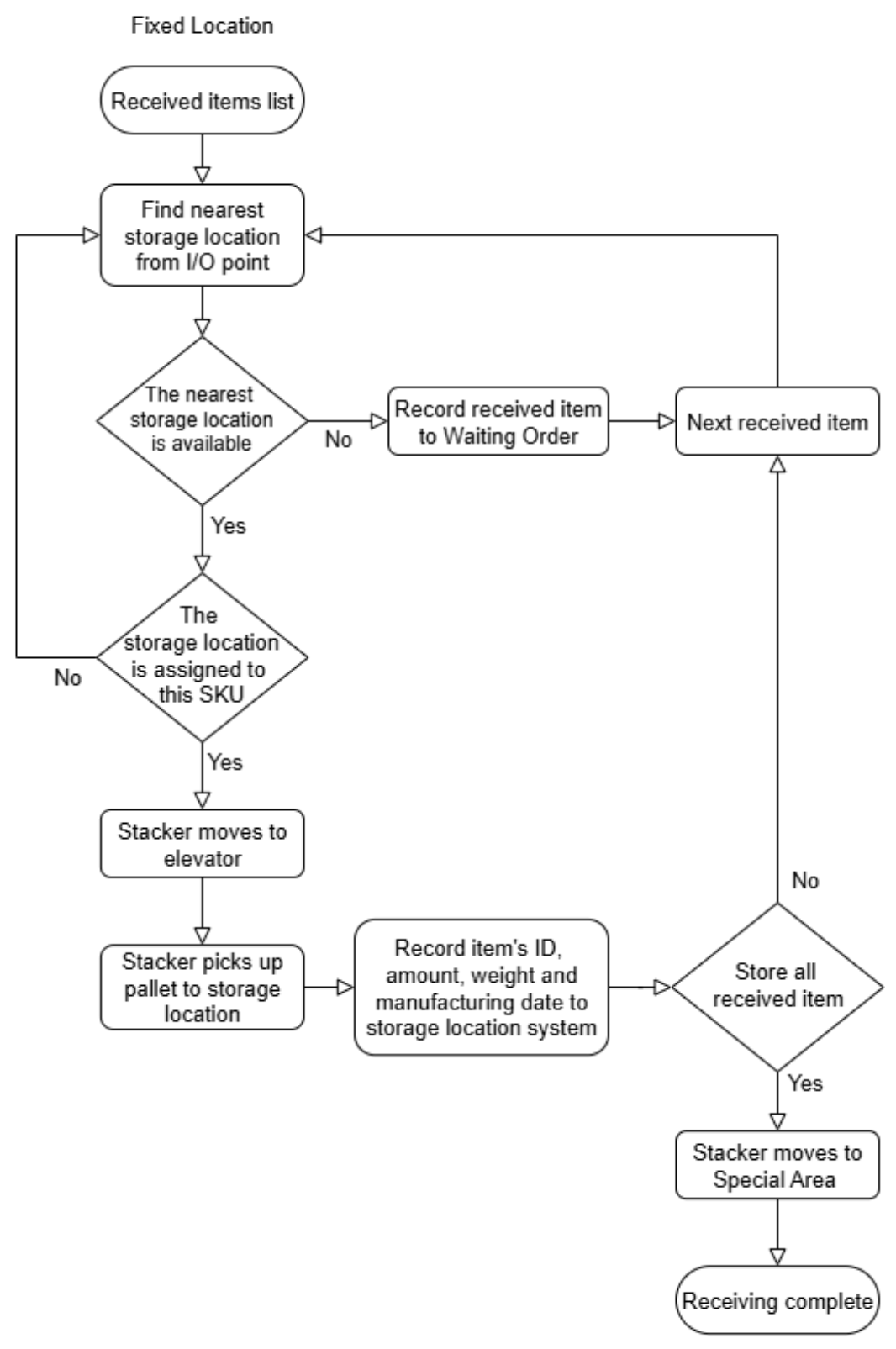

Fig. 7 Illustration of fixed-location storage in the warehouse simulation model.

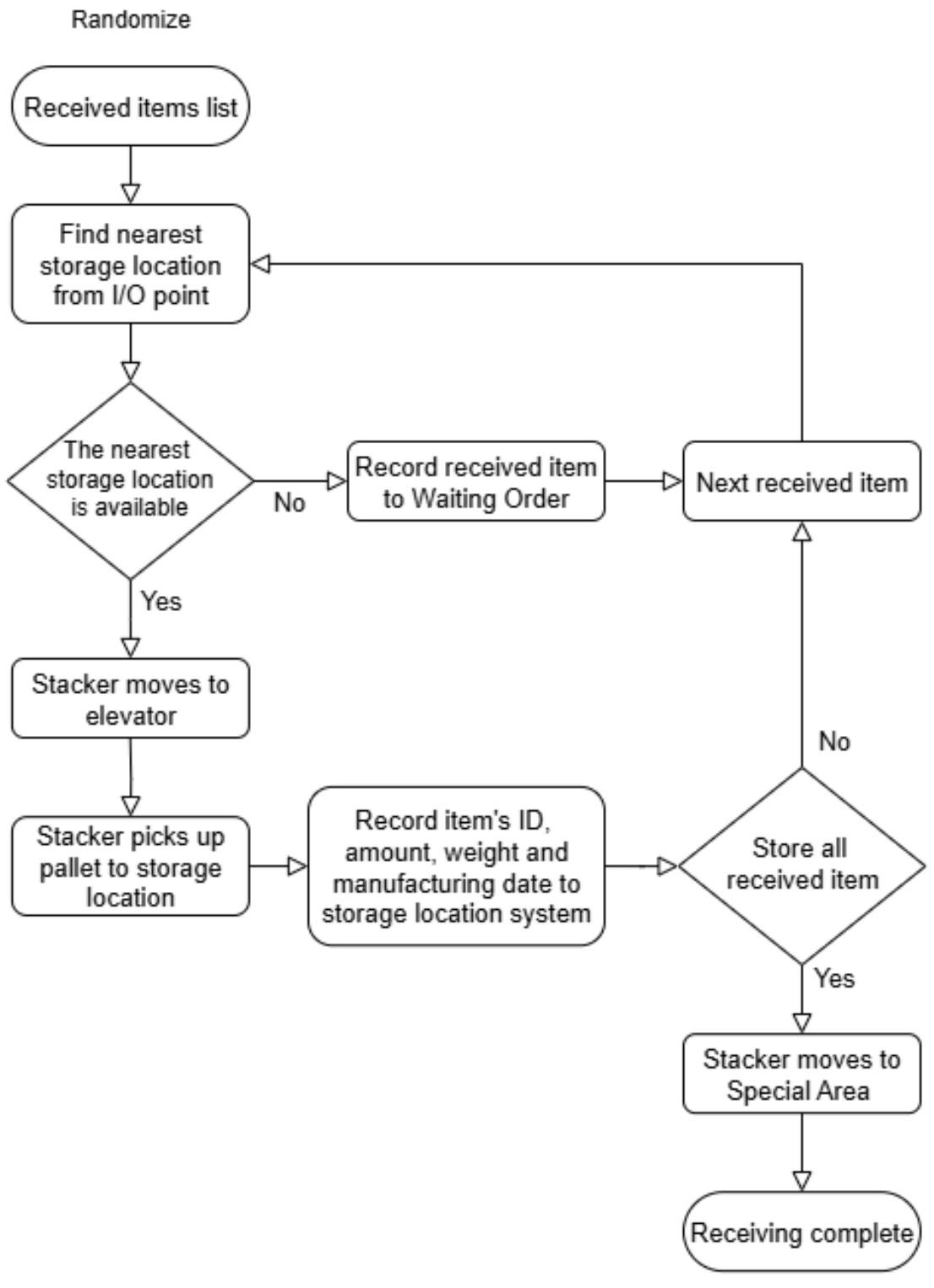

Fig. 8 Illustration of randomized storage in the warehouse simulation model.

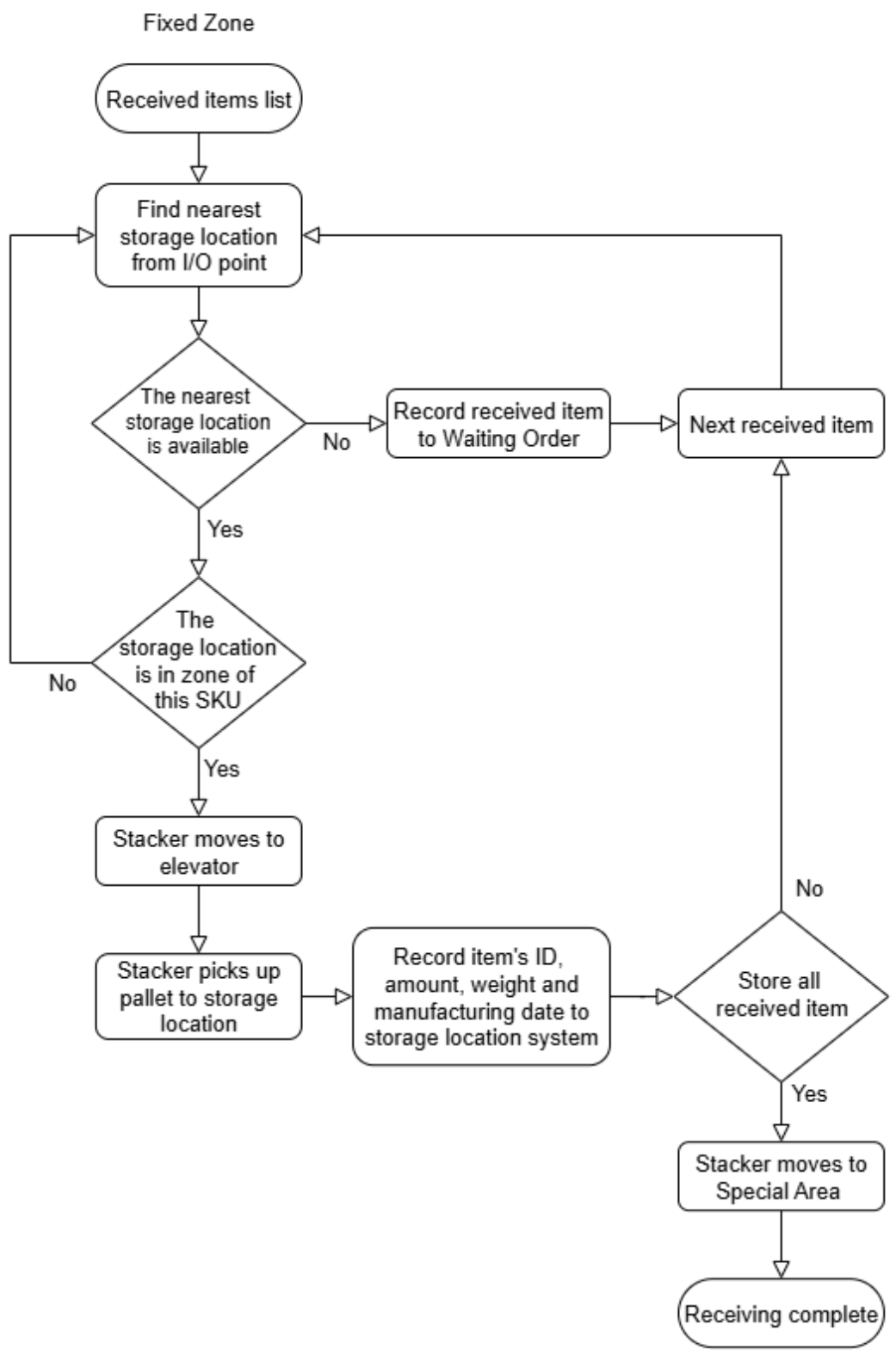

Fig. 9 Illustration of fixed-zone storage in the warehouse simulation model.

## VI. NUMERICAL EXPERIMENT

The simulator models 28,129 order lines over four weeks, covering 153 products across 1,149 storage locations. Orders are picked individually, requiring traversal of the entire storage area. The study focuses on how many slots each product receives, not their exact positions, and treats walking and handling times as constants. Table I lists all simulation variables, with only a subset used for the initial picking-focused scenarios.

Space allocation uses a priority-based algorithm: each product gets one slot, then additional slots are assigned iteratively based on dispatching ratios until all slots are distributed. Two allocation rules create distinct scenario for the simulation. These simulation rules are similar to the set-up scenarios from Gagliardi, Renaud, and A. Ruiz [10].

Scenario 1 (Homogeneous Sharing): Slots are evenly distributed among products.

Scenario 2 (Demand-Based): High-demand products receive more slots based on average picks per product.

Initial tests use only real company data. While statistical demand modeling is ongoing, four weekly experiments were conducted to evaluate the approach, though robustness and variance analysis are not yet possible. The system is terminating, with all products fully replenished weekly, starting from the same initial inventory. Walking speed and replenishment time are kept constant, as recommended by the industrial partner.

The average results in Table 1 show no clear best strategy. Table II's 95% confidence intervals are too wide to draw statistically significant conclusions.

Paired-sample tests show Scenario 2 outperforms Scenario 1, though more replications are needed for reliable results. Location-sharing strategies clearly impact performance, and poorly designed strategies can reduce efficiency. Synchronizing order release with replenishments could further improve operations by reducing stock-outs. Future research will develop proactive replenishment algorithms to anticipate orders and minimize shortages.

## VII. CONCLUSION

This paper presents a discrete-event simulation of warehouse operations, focusing on order-picking and shelf-replenishment. Two storage space allocation rules were tested over eight weeks. Results show that space allocation affects performance, though variability prevents identifying a single best rule, with Rule 2 consistently underperforming. The model is flexible, adaptable to other conveyor-based warehouses, and provides a platform for future studies on system dynamics and the interaction between picking and replenishment.

TABLE I
ALL VARIABLES IN PROCESS SIMULATION

| type | variables | meaning |
|---|---|---|
| Setting | h | numbers of handlift |
| | s | numbers of stacker |
| | oh | numbers of operator/hanlift |
| | os | numbers of operator/stacker |
| | puh | Handlift can put away |
| | pus | Stacker can put away |
| | pah | Handlift can pick up |
| | pas | Stacker can pick up |
| | tfh | Handlift can transfer |
| | tfs | Stacker can transfer |
| | sph | Speed of handlift (cm/sec) |
| | sps | Speed of stacker (cm/sec) |
| | Lsps | Lift Speed of stacker (cm/sec) |
| | tth | Turn Time of handlift (sec) |
| | tts | Turn Time of stacker (sec) |
| Process | BTpu | Base Time (sec) for every pick subprocess |
| | BTpa | Base Time (sec) for every put subprocess |
| | BTs | Base Time (sec) for every sort process |
| | PMpu | Per Master time for master picking process |
| | PMs | Per Master time for master sort process |
| | PPpu | Per Pallet time for pallet picking process |
| | PPpa | Per Pallet time for pallet put process |
| | PPs | Per Pallet time for pallet sort process |
| Plan Duration | PDpu | Pick Up Plan Duration (hour/day) |
| | PDpa | Put Away Plan Duration (hour/day) |
| | PDs | Sort Plan Duration (hour/day) |
| Limit | EAT | Enable Auto Transfer when no more SKU on 1st level of rack |
| | MPW | Max Pallet Weight (kg) |
| | MPV | Max Pallet Volume (%) |
| | WI | Waiting Interval (sec) |
| | OIFW | Operator Info Flow Workload (%) |
| | LR | Lunch/Rest Duration (hr) |
| Simulation | loc | Location ID (Row, Layer, Slot) |
| | Tdx | Total distance of traveling to next node in x axis (cm) |
| | Tdy | Total distance of traveling to next node in y axis (cm) |
| | Tdz | Total distance of traveling to next node in z axis (cm) |
| | sku | Item ERP number |
| | zone | number of type for each sku |
| | weight | weight(kg.) of goods |
| | qty | quantity(pcs.) of goods for each pick demand |
| | shelf_qty | quantity(pcs.) of goods for each location |
| | MFG_Date | Manufacturing date of sku (FIFO) |
| | order_date | date to process of order |
| | order_list | List of Order No. for pick up |
| | receive_list | List of Order No. for put away |
| | start_time | Time to start process |
| | remaining qty | quantity of goods at rack |
| | process_pick_full | time of subprocess full pallet pick (sec) |
| | process_pick_partial | time of subprocess partial pick (sec) |
| | process_put_full | time of subprocess put (sec) |
| | process_put_partial | time of subprocess partial put (sec) |
| | turns_count | number of turns for each equipment |
| | process_move | time of subprocess move (sec) |
| | process_sort_full | time of subprocess full pallet sort (sec) |
| | process_sort_partial | time of subprocess partial sort (sec) |
| | process_waiting | time of subprocess waiting (sec) |

TABLE III
SIMULATION RESULT FROM SCENARIOS

| Week | Scenario 1 | Scenario 2 |
|---|---|---|
| 1 | 103 | 148 |
| 2 | 143 | 150 |
| 3 | 122 | 129 |
| 4 | 97 | 135 |
| Total | 465 | 562 |
| Gap | 0.00% | 20.86% |

TABLE IIII
SIMULATION RESULT FROM SCENARIOS

| | Mean | Lower Limit | Upper Limit |
|---|---|---|---|
| Scenario 1 | 116.25 | 83.19 | 149.31 |
| Scenario 2 | 140.5 | 124.35 | 156.65 |